\documentclass[reprint,amsmath,amssymb, aps, prd,groupedaddress, superscriptaddress, nofootinbib]{revtex4-2}
\usepackage{mathtools}
\usepackage{graphicx}
\usepackage{dcolumn}   % needed for some tables
\usepackage{bm}        % for math
\usepackage{verbatim}   % for math
\usepackage{lineno}
\usepackage{lineno, booktabs, setspace}
\usepackage{wrapfig}
\usepackage{float, array}
\usepackage{amsmath}
\usepackage{amssymb}
\usepackage{verbatim, url}
\usepackage{enumerate}
\usepackage{bm}
\usepackage{notes2bib}
\usepackage{lpic}
\usepackage{pgfplots}
\usepackage[latin1]{inputenc}
\usepackage[extra]{tipa}
\usepackage{accents}
\usepackage{epstopdf}
\usepackage{appendix}
\usepackage{amsthm} 
\usepackage{booktabs}
\newcommand{\ket}[1]{|#1\rangle}

\newcommand{\nn}{\nonumber}
\newcommand{\f}[2] {\frac{#1}{#2}}

\newcommand{\beq}{\begin{equation}}
\newcommand{\eeq}{\end{equation}}
\newcommand{\beqn}{\begin{eqnarray}}
\newcommand{\eeqn}{\end{eqnarray}}

\def\be{\begin{equation}}
	\def\ee{\end{equation}}
\def\bea{\begin{eqnarray}}
	\def\eea{\end{eqnarray}}
\def\ba{\begin{align}}
	\def\ea{\end{align}}

\DeclarePairedDelimiter\abs{\lvert}{\rvert}%
\DeclarePairedDelimiter\norm{\lVert}{\rVert}%
\makeatletter
\let\oldabs\abs
\def\abs{\@ifstar{\oldabs}{\oldabs*}}
\let\oldnorm\norm
\def\norm{\@ifstar{\oldnorm}{\oldnorm*}}
\makeatother

\begin{document}

\title{\boldmath Correlation Functions and Chaotic Behavior of the SYK Chain Model in Pure States}
	
\author{Seyyed M.H. Halataei}
\affiliation{Department of Physics, Shahid Beheshti University, Tehran, Iran}

\date{\today}

\begin{abstract}
	Recent investigations of Rényi entanglement entropy in the SYK chain of Majorana fermions have indicated that the model exhibits slow thermalization when initialized in certain states. The extent to which the heavy modes--believed to underlie this behavior--affect other aspects of the model remains an open question. In this work, I study thermalization and scrambling of information in individual energy eigenstates of the SYK chain using exact diagonalization. I show that two-point correlation functions in finite-energy eigenstates closely match their thermal counterparts and that information scrambling occurs efficiently within these pure states. These results suggest that the slow thermalization observed in entanglement dynamics does not extend to all probes of thermalization and scrambling, even in pure states.

\end{abstract}

\maketitle
%\flushbottom

\section{Introduction}

Chaotic quantum many-body systems are generally expected to thermalize rapidly, even in the absence of external reservoirs. The eigenstate thermalization hypothesis (ETH) provides a framework for understanding this phenomenon \cite{Deutsch1991, Srednicki1994, Srednicki1996, rigol2008thermalization, d2016quantum}. ETH posits that for generic observables, the expectation values in individual energy eigenstates are consistent with thermal averages, implying that a pure state far from equilibrium will evolve such that its observables approach and remain close to thermal values for most of the time.

However, recent investigations of Rényi entanglement entropy in maximally chaotic systems have revealed exceptions to this picture. In particular, a one-dimensional generalization of the Sachdev-Ye-Kitaev (SYK) model--the SYK chain--was found to exhibit slow thermalization when initialized in specific states \cite{Gu2017b, Gu2017}. The role of so-called heavy modes, which are believed to underlie this slow dynamics, and their influence on other observables has been identified as an open question \cite{Gu2017b}.

In a recent companion study \cite{Halataei2025ETH}, I examined ETH in the SYK chain and in two-site SYK models, demonstrating that they satisfy ETH at finite size and thus thermalize rapidly. The present work extends that investigation by analyzing additional probes of thermalization and information scrambling in the SYK chain. Specifically, I study two-point and four-point correlation functions in energy eigenstates, as well as the spectral form factor.

The SYK chain is a one-dimensional lattice model with quenched disorder, in which each site hosts $N$ Majorana fermions subject to random four-fermion interactions \cite{Gu2017}. Neighboring sites are coupled via random bilinear interactions. This model inherits several important features from the original, zero-dimensional SYK model \cite{Kitaev2015, Maldacena2016, sachdev2015bekenstein}, including local criticality, an extensive zero-temperature entropy, and maximal chaos \cite{kitaev2015simple, Maldacena2016, polchinski2016spectrum, maldacena2016bound}. It also displays additional features such as diffusive energy transport and a butterfly velocity that characterizes the spatial spread of chaos \cite{Gu2017}. From a condensed matter perspective, the SYK chain is a rare example of a solvable, strongly interacting, chaotic lattice model--offering a tractable platform for studying thermalization, entanglement propagation, many-body localization, and quantum transport.

From the viewpoint of holography, the SYK chain has been conjectured to be dual to an incoherent black hole \cite{Gu2017}. Since the SYK chain is a closed, unitary system, understanding its thermalization dynamics may yield insights into black hole formation and evaporation, and thereby inform broader discussions surrounding the black hole information paradox \cite{Hawking1976, almheiri2013black}. \footnote{For recent developments in resolving the information paradox, see \cite{Penington2020, Almheiri2019a, Almheiri2020a, Almheiri2019b, Almheiri2020b, Almheiri2020c, Alishahiha2021, Bousso2020, Saad2021}.}

In this work, I examine the spectral form factor, and two- and four-point correlation functions in energy eigenstates of the SYK chain. I perform ensemble averaging over disorder realizations for three different coupling regimes. Using exact diagonalization, I find that these observables in pure states closely resemble their thermal counterparts and that the spectral form factor shows qualitative agreement with predictions from random matrix theory (RMT). These results indicate that the slow thermalization observed in entanglement dynamics \cite{Gu2017b} does not necessarily extend to all probes of thermalization and scrambling, even in pure states. A discussion of the apparent tension with conclusions drawn from Rényi entropy is presented in the final section.

Previous studies have established thermal behavior in two-point functions and scrambling indicators in the complex SYK model, which features all-to-all interactions and no spatial structure \cite{Sonner2017}. The extent to which spatial locality and the absence of all-to-all couplings modify thermalization dynamics remains a key question \cite{Hunter2018}. The present results suggest that, in the analytically tractable regime considered, spatial locality does not obstruct thermalization or scrambling of information in the SYK chain.

The remainder of this paper is organized as follows. Section~\ref{Review} reviews the SYK chain model. Section~\ref{Setup} outlines the numerical implementation, including the construction of the Hamiltonian, the choice of parameter regimes, and the few-body operators used in this study, along with their thermal expectation values. Section~\ref{sec.correlationFunctions} presents the results for two- and four-point functions in energy eigenstates and for the spectral form factor. Finally, Section~\ref{con} summarizes the findings, discusses their implications, and compares them with the entanglement-based study.

\section{Background} \label{Review}
The SYK chain model \cite{Gu2017} is a higher-dimensional generalization of the original Sachdev-Ye-Kitaev (SYK) model \cite{sachdev1993gapless, Kitaev2015, Maldacena2016}. In this section, I first provide a brief overview of the original SYK model before describing the structure and properties of the SYK chain model.

The original SYK model is an ensemble-averaged theory defined by $N$ Majorana fermions with random all-to-all quartic interactions. The Hamiltonian for a single realization of the ensemble is given by
	\beq
	H = \sum_{1\leqslant i < j < k < l \leqslant N} J_{ijkl} \ \chi_i \chi_j \chi_k \chi_l 
	\eeq
	where the Majorana fermion operators satisfy the Clifford algebra
	\beq
	\{ \chi_i, \chi_j \} = \delta_{ij}
	\eeq
	and the couplings ${J_{ijkl}}$ are independent real Gaussian random variables with mean and variance
	\beq
	\overline{J_{ijkl}} = 0, \qquad \qquad	\overline{J_{ijkl}^2} = \f{3! J^2}{N^3}.
	\eeq
	Here, $J$ sets the overall energy scale of the interactions.
	
	The model becomes solvable in the large-$N$ limit and exhibits a number of remarkable features at strong coupling ($N \gg \beta J \gg 1$), including extensive zero-temperature entropy, local quantum criticality characterized by power-law temporal correlations, and maximal chaos. In this regime, the disorder-averaged theory displays emergent conformal symmetry and a duality to nearly-$\mathrm{AdS}_2$ gravity. However, being a $(0+1)$-dimensional model, it lacks spatial structure and therefore cannot realize phenomena such as diffusion or spatially propagating chaos (the butterfly effect).
	
\begin{figure}
			\includegraphics[width=\linewidth]{"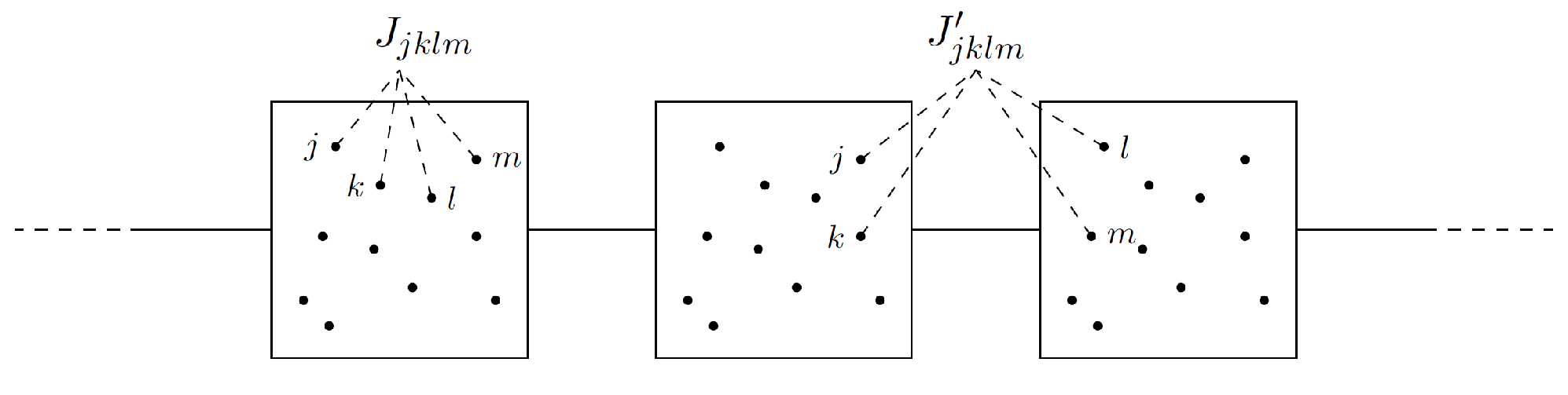"}
			\caption{\label{fig:syk-chain} SYK chain model. The model consists of $M$ sites with periodic boundary conditions. Each site contains $N$ Majorana fermions with on-site SYK interactions. Neighboring sites interact via quartic terms involving two fermions from each site. (Graphic adapted from \cite{Gu2017})}
		\end{figure}

	To introduce spatial locality, Gu et al. \cite{Gu2017} proposed a $(1+1)$-dimensional generalization, known as the SYK chain model. This model consists of a one-dimensional lattice of $M$ SYK sites (also referred to as ``links"), with each site containing $N$ Majorana fermions and interacting with its nearest neighbors via quartic interactions. The structure is illustrated in Fig.~\ref{fig:syk-chain}.
	
In this paper, I focus on the SYK chain model. Like its $(0+1)$-dimensional counterpart, the SYK chain is defined as an ensemble-averaged theory. The Hamiltonian for a single realization takes the form
	\beqn \label{HChain}
	\nn H &=& \sum_{x=1}^M \ \sum_{1 \leqslant i < j < k < l \leqslant N} J_{ijkl,x} \ \chi_{i,x} \chi_{j,x} \chi_{k,x} \chi_{l,x} \\ 
	&+& \sum_{x=1}^M \ \sum_{\substack{ 1 \leqslant i < j \leqslant N \\ 1 \leqslant k < l \leqslant N }} J'_{ijkl,x} \ \chi_{i,x} \chi_{j,x} \chi_{k,x+1} \chi_{l,x+1} 
	\eeqn
	where $J_{ijkl,x}$ and $J'_{ijkl,x}$ are independent zero-mean Gaussian random couplings with variances
	\beq
	\overline{J_{ijkl,x}^2} = \f{3! J_0^2}{N^3}, \qquad \qquad \overline{J^{'2}_{ijkl,x}} = \f{J_1^2}{N^3}.
	\eeq
	Each site $x = 1, \dots, M$ contains $N$ Majorana fermions labeled $\chi_{i,x}$, which satisfy the anticommutation relations
	\beq \label{Cliff}
	\{\chi_{i,x},\chi_{j,y}\}= \delta_{xy} \delta_{ij},
	\eeq
	along with periodic boundary conditions $\chi_{i,0} \equiv \chi_{i,M}$. An effective coupling constant for the model is defined as
	\beq
	J = \sqrt{J_0^2 + J_1^2}.
	\eeq
	In the large-$N$ and strong-coupling limit ($N \gg \beta J \gg 1$), the SYK chain retains many of the intriguing features of the original SYK model, including local criticality, maximal chaos, and an extensive zero-temperature entropy. Owing to its spatial structure, it also exhibits new emergent phenomena such as energy diffusion and a well-defined butterfly effect.
	
	At strong coupling, the SYK chain serves as a solvable example of a strongly correlated, chaotic, and diffusive quantum many-body system. As emphasized in Ref.~\cite{Gu2017}, it provides a valuable platform for exploring diverse aspects of strongly interacting systems, including thermalization, entanglement propagation, and transport.
	
While Ref.~\cite{Gu2017b} examined the growth of R\'enyi entropy and certain aspects of thermalization in the SYK chain model, a direct analysis of correlation functions in energy eigenstates and the spectral form factor remains absent. In the following sections, I investigate whether the correlation functions exhibit thermal behavior and whether the spectral form factor displays features characteristic of random matrix theory (RMT) in the SYK chain model.

\section{Computational Setup} \label{Setup}

I construct the SYK chain model with \( M \) sites, each hosting \( N \) Majorana fermions. The total number of Majorana fermions in the system is thus \( NM \).

In my earlier work~\cite{Halataei2025ETH}, I examined the eigenstate thermalization hypothesis (ETH) for the SYK chain model with \( M = 2 \), \( 3 \), and \( 4 \). Here, I focus on the case \( M = 4 \) to investigate additional indicators of thermalization and quantum information scrambling.

Throughout this work, I set the energy scale \( J = \sqrt{5} \). I consider three parameter choices for the intra-site (\( J_0 \)) and inter-site (\( J_1 \)) coupling strengths:
\begin{enumerate}[(a)]
	\item \( J_0 = 2 \), \( J_1 = 1 \) \hfill (Stronger intra-site coupling)
	\item \( J_0 = J_1 = \sqrt{5/2} \) \hfill (Equal intra- and inter-site coupling)
	\item \( J_0 = 1 \), \( J_1 = 2 \) \hfill (Stronger inter-site coupling)
\end{enumerate}

Even in case (b), where \( J_0 = J_1 \), the model remains distinct from the original SYK model due to the absence of full all-to-all connectivity: the SYK chain includes on-site four-fermion interactions (parameterized by \( J_0 \)) and inter-site interactions restricted to nearest neighbors (parameterized by \( J_1 \)). As a result, this model does not reproduce the original SYK results in the limit \( J_0 = J_1 \).

To build the Hamiltonian in matrix form, I relabel the Majorana fermions using a single index:
\begin{equation}
	\chi_{i,x} \rightarrow \chi_\alpha, \quad \text{with} \quad \alpha = i + (x - 1)N,
\end{equation}
where \( 1 \leq i \leq N \) and \( 1 \leq x \leq M \), so that \( \alpha \in \{1, \ldots, NM\} \). These fermions satisfy the Clifford algebra:
\begin{equation}
	\{ \chi_\alpha, \chi_\beta \} = \delta_{\alpha \beta}.
\end{equation}

Because the Majorana operators are Hermitian (\( \chi_\alpha^\dagger = \chi_\alpha \)), I define complex fermions as follows:
\begin{align}
	c_\alpha &= \chi_{2\alpha} - i \chi_{2\alpha - 1}, \\
	c^\dagger_\alpha &= \chi_{2\alpha} + i \chi_{2\alpha - 1}, \quad \alpha = 1, \ldots, NM/2,
\end{align}
which obey canonical fermionic anticommutation relations:
\begin{equation}
	\{ c_\alpha, c_\beta \} = \{ c^\dagger_\alpha, c^\dagger_\beta \} = 0, \qquad \{ c_\alpha, c^\dagger_\beta \} = \delta_{\alpha \beta}.
\end{equation}

I define a vacuum state \( \ket{0} \) such that
\begin{equation}
	c_\alpha \ket{0} = 0,
\end{equation}
and construct the Hilbert space basis using
\begin{equation}
	(c_1^\dagger)^{n_1} \cdots (c_L^\dagger)^{n_L} \ket{0}, \qquad n_\alpha = 0,1,
\end{equation}
where \( L = NM/2 \). This gives rise to a Hilbert space of dimension \( 2^{NM/2} \).

To obtain the matrix representation of the Majorana operators, I use the recursive construction~\cite{Sarosi2018}:
\begin{align}
	\chi_\beta^{(K)} &= \chi_\beta^{(K-1)} \otimes 
	\begin{pmatrix}
		-1 & 0 \\
		0 & 1
	\end{pmatrix}, \quad \beta = 1, \ldots, 2K - 2, \\
	\chi_{2K - 1}^{(K)} &= I_{2^{K-1}} \otimes 
	\begin{pmatrix}
		0 & 1 \\
		1 & 0
	\end{pmatrix}, \\
	\chi_{2K}^{(K)} &= I_{2^{K-1}} \otimes 
	\begin{pmatrix}
		0 & -i \\
		i & 0
	\end{pmatrix},
\end{align}
where \( I_d \) is the \( d \times d \) identity matrix, and \( K = 1, \ldots, NM/2 \). The recursion is initialized with
\begin{equation}
	\chi_1^{(1)} = 
	\begin{pmatrix}
		0 & -i \\
		i & 0
	\end{pmatrix} = Y, \quad
	\chi_2^{(1)} = 
	\begin{pmatrix}
		0 & 1 \\
		1 & 0
	\end{pmatrix} = X.
\end{equation}

The full set of \( 2^{NM/2} \times 2^{NM/2} \) matrices \( \{\chi_\alpha\} \) is generated in this way. One can then recover the original site and index labels using
\begin{equation}
	i = \alpha \bmod N, \qquad x = \frac{\alpha - i}{N} + 1.
\end{equation}
This enables the explicit construction of the SYK chain Hamiltonian as given in Eq.~\eqref{HChain}.

This construction maps every pair of Majorana fermions (indexed by Latin letters) to a single complex fermion (indexed by Greek letters). In the language of complex fermions, the system comprises $M$ sites, each hosting $N/2$ complex fermions (with $N$ even throughout this work). Specifically, the indices $\alpha = 1, \ldots, N/2$ correspond to site $x = 1$; $\alpha = N/2 + 1, \ldots, N$ to site $x = 2$; $\alpha = N + 1, \ldots, 3N/2$ to site $x = 3$; and so on. The range $\alpha = \frac{N(M - 1)}{2} + 1, \ldots, \frac{NM}{2}$ corresponds to site $x = M$. Expressing operators in terms of complex fermions proves more convenient in the analyses that follow.

In subsequent sections, I examine correlation functions and indicators of chaotic behavior for specific two-body hopping operators. These operators are non-extensive and can be fully non-local--acting between fermions on distant sites. For example, a hopping operator between sites 1 and 3 is defined as
\begin{equation}
	\hat{h}_{13} = c^\dagger_{N+1} c_1 + c^\dagger_1 c_{N+1}.
\end{equation}
Here, \( \alpha = 1 \) is a fermion on site \( x = 1 \), while \( \alpha = N+1 \) is on site \( x = 3 \). This operator thus involves the Majorana modes \( \chi_{1,1}, \chi_{2,1} \) and \( \chi_{1,3}, \chi_{2,3} \).

Similarly, the hopping operator between sites 2 and 4 is defined as
\begin{equation}
	\hat{h}_{24} = c^\dagger_{\frac{3N}{2}+1} c_{\frac{N}{2}+1} + c^\dagger_{\frac{N}{2}+1} c_{\frac{3N}{2}+1}.
\end{equation}
Here, \( \alpha = N/2 + 1 \) corresponds to a fermion on site \( x = 2 \), and \( \alpha = 3N/2 + 1 \) corresponds to one on site \( x = 4 \). This operator involves the Majorana modes \( \chi_{1,2}, \chi_{2,2} \) and \( \chi_{1,4}, \chi_{2,4} \).

Despite their non-local nature, these hopping operators remain non-extensive as they act on only a few fermionic modes.
	
The hopping operators exhibit a notable symmetry: their diagonal matrix elements in the energy eigenbasis vanish identically,
\begin{equation} \label{hopping-diagonal}
	\langle n | \hat{h}_{\alpha \beta} | n \rangle = 0,
\end{equation}
as a direct consequence of the fermionic structure of the model and the Hermiticity of the operators. Accordingly, the microcanonical average also vanishes:
\begin{equation} \label{hopping-micro}
	\langle \hat{h}_{\alpha \beta} \rangle_\text{micro} = 0.
\end{equation}
These identities will be used repeatedly in the following section.

\section{Correlation Functions and Signatures of Quantum Chaos \label{sec.correlationFunctions}}

In a companion work \cite{Halataei2025ETH}, I examined one-point functions of nonextensive operators in the SYK chain model and demonstrated the applicability of the eigenstate thermalization hypothesis (ETH) in that context. The present study extends that analysis by investigating higher-point correlation functions, with the aim of probing dynamical features of the model that are indicative of quantum chaotic behavior. I examine how closely correlation functions evaluated in pure quantum states approximate their thermal counterparts, drawing upon both numerical simulations and analytic arguments.

While many of the correlation functions considered here have been previously analyzed in the thermal ensemble \cite{cotler2017black, davison2017thermoelectric, Garcia2018, kitaev2015simple, Maldacena2016, maldacena2016bound, fu2016numerical}, my emphasis is on their behavior in pure states, particularly in individual energy eigenstates. From the perspective of the holographic correspondence, this translates into studying how accurately correlation functions computed in an incoherent black-hole background are approximated by correlations in pure states.

The structure and methodology of this section are inspired by Ref. \cite{Sonner2017}, which investigated similar questions in the 0D complex SYK model. Here, I extend those ideas to the spatially extended SYK chain to explore how thermalization and scrambling of information manifest in the presence of locality.

I begin with the spectral form factor, which serves as a diagnostic of chaotic dynamics. In contrast to other observables, the spectral form factor does not distinguish between pure and mixed states, as it is naturally defined in terms of an analytically continued partition function. Alternatively, it may be interpreted as the fidelity of a suitably chosen pure state \cite{del2017scrambling}.

\subsection{The Spectral Form Factor \label{sec:specFormFac}}

The spectral form factor is a widely studied diagnostic in random matrix theory (RMT) \cite{brezin1997spectral}, offering a sensitive probe of spectral properties such as level repulsion, eigenvalue correlations, and the discreteness of the energy spectrum. Its behavior at late times is particularly informative, as it reflects universal signatures of quantum chaos. The spectral form factor is most naturally defined via the analytically continued thermal partition function: 
\be
{\cal S}(\beta, t) := \frac{Z(\beta + i t) Z(\beta - it)}{Z(\beta)^2}\,.
\ee

\begin{figure}[t!]
	\begin{center}
		\includegraphics[width=0.49\textwidth]{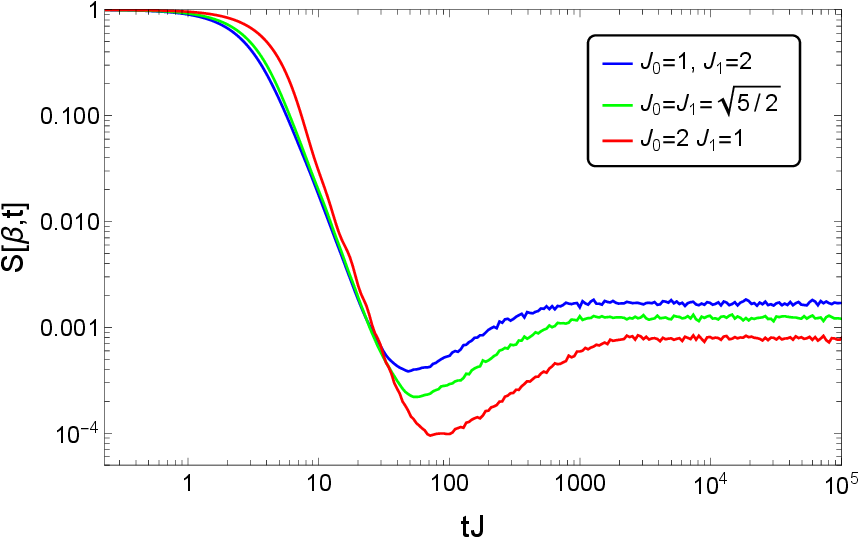} 
		\caption{\small Spectral form factor computed from 1000 disorder realizations for three distinct sets of coupling parameters at fixed system size $N M= 24$,  with inverse temperature $\beta = 1.5$. In all cases, the plot exhibits the characteristic early-time decay, linear ramp, and saturation plateau associated with chaotic spectra.
		\label{fig:SYKsff}}
	\end{center}
\end{figure}
\noindent Figure~\ref{fig:SYKsff} displays ${\cal S}(\beta, t)$ as a function of time for the SYK chain, averaged over 1000 disorder realizations and evaluated at inverse temperature $\beta = 1.5$. The data reveals the familiar three-stage structure also observed in RMT: an initial decay (the ``dip''), followed by a linear ramp, and eventual saturation at a plateau \cite{cotler2017black, del2017scrambling}. While this qualitative structure resembles RMT behavior, the quantitative features--such as the dip time and plateau time--depend on model-specific details \cite{cotler2017black}.

Notably, we observe that both the dip depth and the timescale for ramp onset decrease as the ratio $J_0/J_1$ becomes smaller. This trend is consistent with the interpretation that, in the limit $J_0/J_1 \to 0$, the system deviates from SYK-like behavior and no longer exhibits features characteristic of random matrix theory.

\subsection{Two-Point Correlation Function in Energy Eigenstates}

I now turn to the study of correlation functions involving non-extensive operators. Specifically, I focus on the two-site hopping operator \( h_{13} \), introduced in Sec.~\ref{Setup}. Although I select the first and third sites and particular Majorana fermions on each, any such choice of site pair and fermion indices can be considered.

To analyze the behavior of two-point functions in energy eigenstates, I consider the three coupling configurations defined previously: \( J_0 = 2, J_1 = 1 \); \( J_0 = J_1 = \sqrt{5/2} \); and \( J_0 = 1, J_1 = 2 \). For notational clarity, I denote the Hermitian hopping operator \( h_{13} \) by \( \mathcal{O} \). The time-dependent two-point function in an eigenstate \( |n\rangle \), with energy \( E_n \), is defined as
\beq
G^n(t) = \langle \mathcal{O}(t)\mathcal{O} \rangle_{E_n} := \langle n|e^{iHt} \mathcal{O} e^{-iHt} \mathcal{O}|n\rangle.
\eeq
Owing to the vanishing of the diagonal matrix elements of \( \mathcal{O} \) [cf. Eq.~\eqref{hopping-diagonal}], the connected and full two-point functions coincide:
\beq
G^n_c(t) = G^n(t) - \langle \mathcal{O} \rangle_{E_n}^2 = G^n(t).
\eeq

\begin{figure}[h!]
	\centering
	\includegraphics[width=0.49\textwidth]{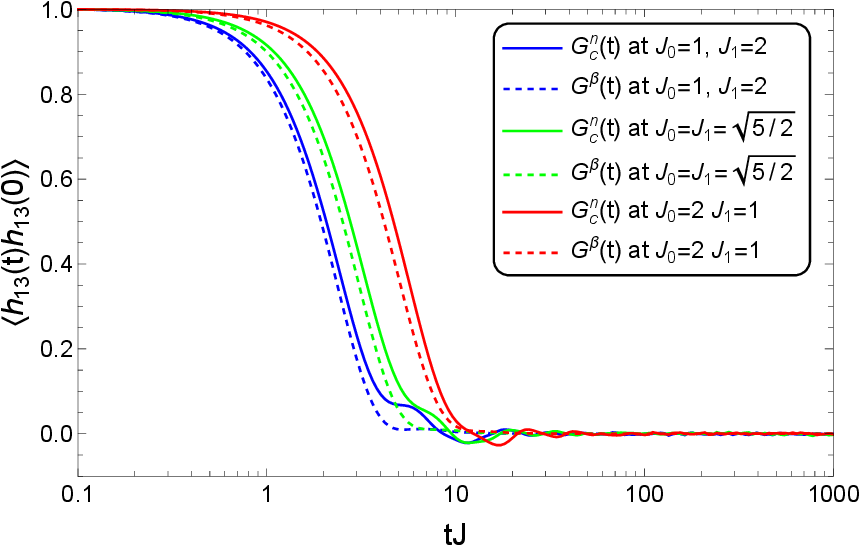} 
	\caption{\small Two-point correlation functions of the hopping operator between two sites of the SYK chain with $NM = 24$ Majorana fermions.  Each correlation function is averaged over $1000$ realizations of the disorder. The eigenstate correlators \( G^n_c(t) \) and the canonical correlators \( G^\beta(t) \) exhibit excellent agreement across all three choices of intra- and inter-site coupling strengths.}
	\label{fig:SYK2ptI}
\end{figure}

Numerical evaluation reveals that \( G^n(t) \) exhibits a rapid decay followed by persistent fluctuations about zero, as illustrated in Fig.~\ref{fig:SYK2ptI}. The time average of this correlator is readily computed:
\beq
\overline{G^n(t)} = \lim_{T \to \infty} \frac{1}{T} \int_0^T \sum_m e^{i(E_n - E_m)t} |\mathcal{O}_{nm}|^2 dt = |\mathcal{O}_{nn}|^2.
\eeq
In accordance with the eigenstate thermalization hypothesis (ETH), these late-time fluctuations are suppressed by a factor of \( \sim e^{-S/2} \), where \( S \) is the thermodynamic entropy at energy \( E_n \).

ETH further implies that two-point functions in individual eigenstates should closely approximate their microcanonical counterparts in the thermodynamic limit. This expectation is borne out numerically: the long-time average of \( G^n(t) \) matches the microcanonical average defined in Eqs.~(\ref{hopping-diagonal})--(\ref{hopping-micro}). Figure~\ref{fig:SYK2ptI} provides empirical support for this correspondence. These observations are perhaps not surprising given that disorder-averaging over couplings plays a role analogous to microcanonical ensemble averaging.

To explore thermalization more broadly, I now compare eigenstate correlators to their canonical ensemble analogues. Let \( \rho = e^{-\beta H} / Z(\beta) \) denote the canonical density matrix at inverse temperature \( \beta \), with energy \( E(\beta) \) determined by
\beq \label{eq:E_beta}
E(\beta) = \f{1}{Z} \text{Tr}[e^{-\beta H} H ].
\eeq

The canonical two-point function is given by
\beq
G^\beta(t) = \frac{1}{Z(\beta)} \text{Tr}[e^{-\beta H} \mathcal{O}(t)\mathcal{O}],
\eeq
and, by the same reasoning as above, it coincides with its connected form 
\beq
G^\beta_c(t) = G^\beta(t) - \left( \frac{1}{Z(\beta)} \text{Tr}[e^{-\beta H} \mathcal{O}] \right)^2.
\eeq

Across all parameter choices, these thermal correlators display the expected features: an initial exponential decay at early times, a subsequent power-law decay at intermediate times, and saturation to a plateau at late times.

Figure~\ref{fig:SYK2ptI} compares \( G^n_c(t) \) to the corresponding thermal correlator \( G^\beta(t) \) across different coupling regimes. To align the energies of the eigenstate and thermal ensemble, I fix the inverse temperature at \( \beta = 1.5 \) and select the eigenstate \( \ket{n} \) whose energy \( E_n \) most closely matches the thermal average energy \( E(\beta) \). The results show excellent agreement at both early and late times. Small deviations appear at intermediate times, particularly near the end of the power-law decay regime, where \( G^n_c(t) \) exhibits residual oscillations around its long-time plateau, whereas \( G^\beta(t) \) approaches the plateau more smoothly. These discrepancies diminish with increasing Hilbert space dimension and are expected to vanish in the thermodynamic limit \( N \to \infty \), which is of particular relevance to the conjectured gravitational dual, as it corresponds to the semiclassical regime.

In summary, the two-point function in individual energy eigenstates of the disorder-averaged theory exhibits behavior consistent with thermalization, most closely aligning with microcanonical predictions. Minor distinctions between canonical and eigenstate correlators are evident at finite sizes but are expected to disappear in the large-\( N \) limit.

\subsection{Four-Point Correlation Function in Energy Eigenstates} \label{sec.OTOC}

\begin{figure}[h!]
	\centering
	\includegraphics[width=0.49\textwidth]{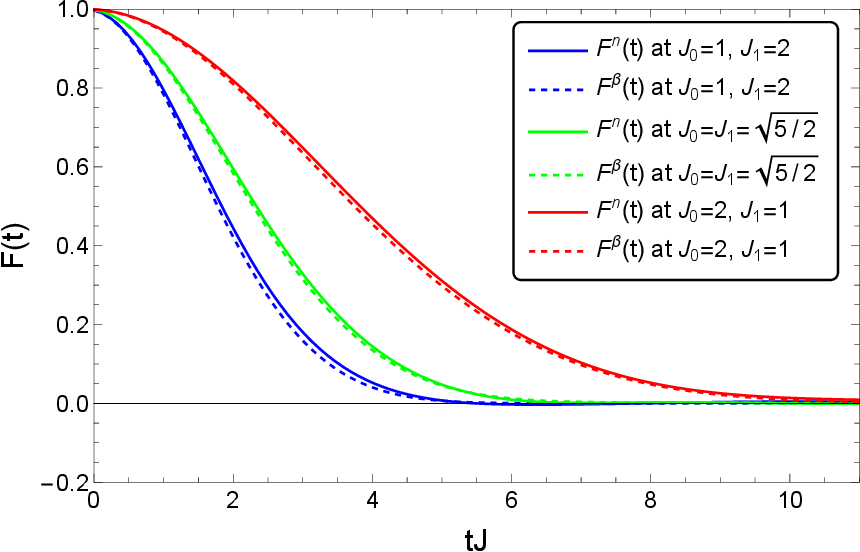}
	\caption{\small Four-point out-of-time-order correlation function (OTOC) of the hopping operators in the SYK chain with $NM = 24$ Majorana fermions. Each correlation function is averaged over $30$ disorder realizations. The eigenstate $\ket{n}$ is selected such that its energy $E_n$ matches the average energy of the thermal ensemble at inverse temperature $\beta = 1.5$. The eigenstate OTOCs \( \mathcal{F}^n(t) \) and their thermal counterparts \( \mathcal{F}^\beta(t) \) exhibit excellent agreement across all three choices of intra- and inter-site coupling strengths.}
	\label{fig:SYK4ptI}
\end{figure}

I now examine the four-point correlation function, which probes early-time quantum chaos through the Lyapunov exponent \( \lambda_L \). In chaotic systems, an appropriately defined out-of-time-order correlator (OTOC) typically exhibits an initial exponential decay up to the scrambling time. For a thermal ensemble, this behavior is captured by
\begin{equation}
	\langle W(t)V(0) W(t) V(0) \rangle_\beta \approx 1 - \alpha e^{\lambda_L t},
\end{equation}
where \( \alpha \) is a constant and the expectation value is taken with respect to the thermal state. This relation defines a quantum analog of the classical Lyapunov exponent.

In the SYK chain, the Lyapunov exponent in the thermal regime is known to saturate the universal upper bound \( \lambda_L = 2\pi/\beta \) \cite{maldacena2016bound, Gu2017}. The aim of this section is to test whether such behavior persists at the level of individual energy eigenstates. Specifically, I consider the correlator
\begin{equation} \label{eq.EstateLyapunov}
	\langle n| W(t)V(0) W(t) V(0) |n\rangle,
\end{equation}
and investigate whether it displays a similar exponential decay at early times.

To align the energy of the eigenstate with that of the thermal ensemble, I fix the inverse temperature to \( \beta = 1.5 \) and select the eigenstate \( \ket{n} \) whose energy \( E_n \) is closest to the thermal average energy,
\begin{equation}
	E(\beta) = \frac{1}{Z} \text{Tr}[e^{-\beta H} H],
\end{equation}
where \( Z = \text{Tr}[e^{-\beta H}] \) denotes the partition function.

While the properties of thermal OTOCs have been extensively studied in prior work \cite{maldacena2016bound, Kitaev2015, fu2016numerical, Maldacena2016, Gu2017}, the present analysis focuses on their behavior within individual eigenstates, aiming to determine whether the same chaotic signatures extend beyond ensemble averages.

As in previous sections, I work with the two-site hopping operators \( \hat{h}_{\alpha\beta} \). For the SYK chain, I define
\begin{equation}
	W = \hat{h}_{13}, \qquad V = \hat{h}_{24},
\end{equation}
and evaluate the normalized four-point function
\begin{equation}
	\mathcal{F}(t) = \frac{\left[ \langle W(t) V(0) W(t) V(0) \rangle + \langle V(t) W(0) V(t) W(0) \rangle \right]} {\langle W(0) W(0) V(0) V(0) \rangle},
\end{equation}
with the expectation values computed either in a thermal ensemble (\( \mathcal{F}^\beta(t) \)) or in an individual eigenstate (\( \mathcal{F}^n(t) \)).

Before scrambling sets in, this function is expected to take the form
\begin{equation} \label{eq.OTOCScramble}
	\mathcal{F}(t) = \mathcal{F}_0 - \alpha e^{\lambda t},
\end{equation}
where \( \lambda \) characterizes the rate of exponential decay. In the large-\( N \) limit, the SYK chain is known to saturate the chaos bound,
\begin{equation}
	\lambda = \frac{2\pi}{\beta},
\end{equation}
as established in Ref.~\cite{Gu2017}.

As shown in Fig.~\ref{fig:SYK4ptI}, my numerical results demonstrate that the eigenstate OTOCs closely track their thermal counterparts across all three sets of intra- and inter-site couplings: $J_0 = 2$, $J_1 = 1$; $J_0 = J_1 = \sqrt{5/2}$; and $J_0 = 1$, $J_1 = 2$. Notably, the exponential decay rate and the scrambling time are in excellent agreement. These findings support the conclusion that eigenstate thermalization extends beyond two-point observables and governs even higher-order correlators that are sensitive to quantum chaos. Therefore, in the large-\( N \) limit, the eigenstate OTOCs \( \mathcal{F}^n(t) \) are expected to exhibit exponential growth with a Lyapunov exponent consistent with the thermal value.

\section{Discussion and Conclusion} \label{con}

In this study, which complements the analysis presented in Ref.~\cite{Halataei2025ETH}, I explored the spectral form factor and pure-state correlation functions in the strongly coupled, spatially extended (1+1)-dimensional SYK chain model. Through exact diagonalization and averaging over a large ensemble of disorder realizations, I examined the behavior of two- and four-point functions as well as the spectral form factor. The primary objective was to determine whether the SYK chain exhibits features of quantum chaos typically associated with random matrix theory (RMT), and whether individual energy eigenstates manifest thermal behavior.

The main findings can be summarized as follows. First, the spectral form factor exhibits hallmark features of RMT and quantum chaotic systems, including the characteristic dip-ramp-plateau structure. Second, two- and four-point correlation functions evaluated in individual energy eigenstates closely match their thermal counterparts. These thermal correlation functions, together with the spectral form factor, in turn reveal signatures of rapid thermalization, efficient information scrambling, and quantum chaos. Thus, the SYK chain, despite its spatial locality and lack of all-to-all couplings, displays dynamical behavior consistent with strongly chaotic quantum systems.

A key conclusion is that thermalization--when probed through correlation functions--occurs significantly more rapidly than previously inferred from studies based on R\'enyi entropies~\cite{Gu2017b}. This discrepancy may arise from the different observables used to probe thermalization. In particular, in the strong-coupling regime of the SYK chain ($N \gg \beta J \gg 1$), the dominant low-energy dynamics is governed by collective soft modes associated with time reparametrization. These modes appear to equilibrate few-body operators efficiently, even in the presence of local interactions. In contrast, the heavier modes identified in Ref.~\cite{Gu2017b}--which were argued to govern the slow growth of R\'enyi entropies--do not appear to play a significant role in the behavior of correlation functions, the spectral form factor, or the onset of scrambling.

Further support for this interpretation comes from the analysis in Ref.~\cite{Sohal2022}, which demonstrates that thermalization in the SYK chain is strongly state-dependent. Specifically, states with sufficiently high effective temperature thermalize quickly, even when entanglement-based diagnostics are used. That work also provides evidence that the subthermal behavior reported earlier is likely an artifact of the large-$N$ limit, rather than a generic feature at finite $N$. These observations are consistent with the present findings and reinforce the conclusion that thermalization and information scrambling, when measured via correlation functions, occur on comparatively short timescales.

The implications of these results are twofold. First, they confirm that spatial locality in the SYK chain does not inhibit the emergence of chaotic dynamics and thermal behavior, highlighting that fast thermalization and scrambling are not exclusive to zero-dimensional models with all-to-all interactions. Second, from a holographic standpoint, where the SYK chain has been conjectured to correspond to an incoherent black hole, these results suggest that the dual gravitational system likewise exhibits rapid thermalization and scrambling of few-body observables--supporting the broader picture of fast thermalization in holographic systems.

This work also extends insights obtained in a zero-dimensional SYK-like model~\cite{Sonner2017} to higher-dimensional systems with spatial structure. The findings indicate that the essential mechanisms underlying chaos, thermalization, and scrambling--particularly those tied to collective low-energy dynamics--remain effective in systems with local interactions. This suggests a degree of universality in the thermal behavior of strongly interacting quantum chaotic systems.

Several directions for future research naturally follow. While this study focused on a specific operator--the two-site hopping operator--it would be valuable to investigate whether similar behavior extends to more general few-body observables. Moreover, applying the same methodology to higher-dimensional generalizations of the SYK, such as those discussed in Refs.~\cite{Berkooz2017,Gu2017}, may help establish the robustness and universality of the present results across a wider class of strongly coupled models.

\bigskip

In conclusion, this work demonstrates that the SYK chain model, despite lacking all-to-all connectivity and despite previous suggestions of slow entanglement dynamics, thermalizes rapidly, scrambles information effectively, and exhibits hallmark features of quantum chaos. These results underscore the resilience of thermalization and information scrambling in strongly correlated, spatially local quantum systems.

	%****************************************** Acknowledgement ********************************************************
	
\section{acknowledgments}
	I would like to thank Mohsen Alishahiha and Akbar
	Jafari for their special inspiration and insightful discussions. I am also grateful to Mark Srednicki, Manuel Vielma,
	Julian Sonner, and Mahdieh Piranaghl for many fruitful
	conversations. I especially acknowledge Junyu Lin for his
	invaluable assistance with the code during the early stages
	of this work.

	%********************************************* Appendix

\bibliographystyle{unsrt}

\bibliography{allrefs}
%\bibliography{../../Bibliography/allrefs} 
\end{document}